\documentclass[eqsecnum,preprintnumbers,showpacs,showkeys,aps, 10pt]{revtex4}
\usepackage{amsmath, amssymb, bm}

\usepackage{hyperref}
\usepackage{graphicx}
\newcommand*{\ket}[1]{\ensuremath{|#1\rangle}}

\begin{document}

\title{Finer Characterizations of Pure Bipartite Entanglement }

\author{Che-Hsu Li}\email{sunine1130@gmail.com}

\affiliation{Department of Physics, National Cheng Kung University, Taiwan}

\begin{abstract}

A new criterion necessary and sufficient for the separability of pure bipartite systems for arbitrary finite dimensions is demonstrated; and the corresponding finer quantitative measures or characterizations of entanglement (beyond mere separability or non-separability determination) are discussed. Based on this criterion, we proved that the well-known Peres-Horodecki positivity-of-partial-transform criterion is also necessary and sufficient for separability in the case of pure bipartite systems. The maximum value of entanglement, and the corresponding maximally-entangled states are also worked out in detail.

\end{abstract}



\maketitle

\section{Introduction}

Entanglement is one of the most remarkable traits manifested by quantum systems. It was precisely the puzzling property involved in the discussion by Einstein-Podolsky-Rosen in their well-known article in 1935\cite{EPR} to expose the ``incompleteness" of quantum mechanics.  It is also the very characteristic regarded by E. Schr\"odinger as the crucial feature which distinguishes quantum from classical systems\cite{SHRO}. Attempts to mimic and ``explain" probabilistic quantum mechanics by local realistic hidden-variable theories suffered a serious setback when J.S. Bell demonstrated that, barring  super-luminal communications and/or other further assumptions, violations of a certain inequality (now named Bell Inequality) in entangled systems cannot be reproduced by hidden variable theories. In addition he suggested experimental verifications of these violations in the real world\cite{BELL}. Such an experiment, carried out by A. Aspect et al. in 1981\cite{BELLTEST}, produced evidence of violations of Bell's inequality which at the same time obey the predictions of quantum mechanics. J. F. Clauser, M. A. Horne, A. Shimony, and R. A. Holt (CHSH)\cite{CHSH} subsequently formulated a more convenient version of the Bell inequality which is more often discussed. Entanglement has emerged as a key resource in the most important applications of Quantum Information Science, such as quantum teleportation\cite{Tel} and quantum cryptography\cite{crypt}.

In the early days, considerations of entanglement was mostly qualitative. Bell's work can in fact  be regarded as an early attempt to quantify entanglement, and over the years various entanglement measures have been proposed, such as the entropy of entanglement\cite{ENTR}, the maximum expectation value of the CHSH operator, as well as other quantitative measures\cite{INTR}. In this work, which is based on\cite{THES}, we propose a new necessary and sufficient criterion for separability for pure finite-dimensional bipartite systems which corresponds, in general, to a group of entanglement measures. This group of measures uses multiple parameters, rather than only one value, to characterize and quantify entanglement beyond just mere separability (entangled) or non-separability (non-entangled) criterion. The total entanglement is zero if and only if (iff) each and every one of the entanglement parameters vanishes. But the same value of total entanglement can correspond to different combinations of the different entanglement parameters. Thus a finer quantitative characterization of entanglement is realized. Moreover, each entanglement  parameter can be interpreted as the entanglement of a particular qubit-qubit subsytem, and thus lends itself to possible measurement and verification, for instance, via the expectation value of the CHSH operator.

Our formalism focuses on $n \times m$ pure bipartite systems. We decompose the matrix of the total quantum state coefficients into  $2\times2$ sub-matrices. It can then be proven that the total compound system is separable iff all these sub-matrices are separable. This enables us to first establish the {\it necessary and sufficient criterion for separability}, and {\it furthermore} to {\it construct explicit quantitative measures of entanglement which go beyond mere separability or non-separability determination}. The entanglement of a bipartite qubit-qubit system is monotonic to the maximum of the expectation value of the Clauser-Horne-Shimony-Holt(CHSH) operator. This implies that the entanglement parameters defined from our criterion can also be measured by CHSH measurements on each $2\times2$ submatrix which corresponds to a qubit-qubit subsystem. Our results further indicate that beyond qubit-qubit bipartite systems more than one measure is necessary to completely characterize entanglement, and we provide such a set of finer characterizations. The relation between our new criterion and the Peres-Horodecki positivity of partial transform (PPT) criterion\cite{PER}\cite{HOR} for separability is also clarified; and it is demonstrated, using our methods, that PPT is necessary and sufficient for separability of pure finite-dimensional bipartite systems. The maximum value of total entanglement is computed, and the explicit corresponding states with maximal entanglement are worked out.

\section{New criterion for entanglement}\label{sec3}

We focus on generic  $n \times m$ pure bipartite systems $C_{iJ} \ket i \otimes \ket J $ described by  the matrix of state coefficients $C_{iJ}$
\begin{equation}
{\rm state\,\, coefficient\,\, matrix}
\,\,\,\,\,C\,\,\,\,\,=
\left( \begin{array}{ccccc}
C_{11}&C_{12}&C_{13}&...&C_{1m}\\
C_{21}&C_{22}&C_{23}&...&C_{2m} \\
C_{31}&C_{32}&C_{33}&...&C_{3m}\\
\vdots& ...& ...&\ddots& \vdots \\
C_{n1}&C_{n2}&C_{n3}&...&C_{nm}
\end{array}\right).
\end{equation}

\subsection{Definition of submatrix}

We define $2 \times 2$ submatrices of an $n \times m$ matrix $C_{iJ}$ by 
\\\\
\textbf{Definition 1}: 
$$
S^{(a,b)} (C) \equiv
\left( \begin{array}{cc}
C_{a,b}&C_{a,b+1}\\
C_{a+1,b}&C_{a+1,b+1}
\end{array}\right),
\,\,\,\,\,\,\,where\,\, a=1,...,(n-1)\, , \,b=1,...,(m-1).
$$
For an  $n \times m$ matrix, there are $ (n-1) \times(m-1)$ submatrices.
\\\\
\textbf{Theorem 1:}
If $C_{iJ}$ contains no elements equal to zero, $\sum_{a,b} | \det ( S^{(a,b)}) |  = 0   $ is a necessary and sufficient condition for $C_{iJ}$ to be separable.
\\\\
\textbf{Proof:}

First of all,
\begin{equation}
 | \det ( S^{(a,b)}) |  = 0\iff  \det ( S^{(a,b)})   = 0.
\end{equation}
Moreover, from \textbf{Definition 1},
\begin{equation}
 \det ( S^{(a,b)})   = 0 \iff C_{a,b}C_{a+1,b+1}=C_{a,b+1}C_{a+1,b}.
\end{equation}
Since $C_{iJ}$  has no zero elements,
\begin{equation}
C_{a,b}C_{a+1,b+1}=C_{a,b+1}C_{a+1,b} \iff  \frac{ C_{a,b}}{C_{a,b+1}}= \frac{ C_{a+1,b}}{C_{a+1,b+1}}\iff C_{a,b}:C_{a,b+1}=C_{a+1,b}:C_{a+1,b+1}.
\end{equation}
Thus we confirm (if $C_{iJ}$ has no zero elements),
\begin{equation}
 | \det ( S^{(a,b)}) |  = 0 \iff  C_{a,b}:C_{a,b+1}=C_{a+1,b}:C_{a+1,b+1}\label{eq2}.
\end{equation}
Note also that the sum $\sum_{a,b} | \det ( S^{(a,b)}) |  = 0  \iff all \,\,\, | \det ( S^{(a,b)}) |  = 0  $.
Therefore, if  $C_{iJ} $ satisfies  $\sum_{a,b} | \det ( S^{(a,b)}) |  = 0$, then
\begin{equation}
C_{11}:C_{12}=C_{21}:C_{22}=C_{31}:C_{32}=...=C_{n1}:C_{n2}.
\end{equation}
This is equivalent to
\begin{equation}
C_{11}:C_{21}:C_{31}:...:C_{n1}=C_{12}:C_{22}:C_{32}:...:C_{n2}.
\end{equation}
Similarly, we have
\begin{equation}
C_{12}:C_{22}:C_{32}:...:C_{n2}=C_{13}:C_{23}:C_{33}:...:C_{n3}.
\end{equation}
Using the the same arguments, we arrive finally at
\begin{subequations}
\begin{align}
& C_{11}:C_{21}:C_{31}:...:C_{n1}\\
=&C_{12}:C_{22}:C_{32}:...:C_{n2}\\
=&C_{13}:C_{23}:C_{33}:...:C_{n3}\\
& \vdots  \notag  \\
=&C_{1m}:C_{2m}:C_{3m}:...:C_{nm}.
\end{align}\label{eq3}
\end{subequations}
This is equivalent to $C_{iJ}$ being separable because the rows  are all respectively proportional to one another and so are the columns. We may choose $(C_{11},C_{21},C_{31},...,C_{n1})$ as the coefficients of $|a \rangle$ and $\frac{(C_{11},C_{12},C_{13},...,C_{1m})}{C_{11}}$ as those of $|b \rangle$ thus yielding
 $|c \rangle =|a \rangle \otimes |b \rangle$ wherein $|c \rangle$ is the state vector corresponding to matrix $C_{iJ}= a_i b_J$.

Conversely, from Eq. (\ref{eq2}) we infer that Eq. (\ref{eq3}) implies  $\sum_{a,b} | \det ( S^{(a,b)}) |  = 0   $. Therefore we can conclude that  $\sum_{a,b} | \det ( S^{(a,b)}) |  = 0   $ is a necessary and sufficient condition for $C$ to be separable (under the premise that $C$ has no zero elements). $\Box $

To extend the results to the case when $C_{iJ}$ may possess vanishing elements, we define the reduced matrix $R$ of a matrix $C$.
\\\\
\textbf{Definition 2}: The reduced matrix $R$ of a larger matrix $C$ is the matrix obtained by eliminating all rows and columns which contain all zero elements.

For instance, if $C$ is
$$
\left( \begin{array}{ccc}
a&0&b\\
c&0&d\\
e&0&f
\end{array}\right),
$$
then its reduced matrix $R$ is
$$
\left( \begin{array}{ccc}
a&b\\
c&d\\
e&f
\end{array}\right);
$$
if $C$ is
$$
\left( \begin{array}{ccccc}
a&b&0&0&c\\
d&e&0&0&f\\
0&0&0&0&0\\
g&h&0&0&i
\end{array}\right),
$$
then its reduced matrix $R$ is
$$
\left( \begin{array}{ccc}
a&b&c\\
d&e&f\\
g&h&i
\end{array}\right).
$$

Obviously,
\begin{equation}
C\,\, is \,\,separable \iff \,\, R\, \,is\,\, separable. \label{eq4}
\end{equation}
\\\\
\textbf{Theorem 2}: If a reduced matrix $R$ has at least one zero, then it must be entangled.
\\\\
\textbf{Proof}:

Suppose $R$ has a zero in $k^*$-th row and $L^*$-th column, i.e. $R_{k^*L^*}=0$. If we also {\it assume} $R$ is separable, namely $R_{iJ}=a_ib_J \quad \forall i, J$.
We may illustrate this situation by
$$
\bordermatrix{&b_1&b_2&...&b_{L^*}&... \cr
  a_1 &   &  &   & \vdots  &   \cr
    a_2 & &   & & \vdots  \cr
     \vdots &  && & \vdots  \cr
    a_{k^*} &... & ... & ...  &R_{k^*L^*}&...  \cr
  \vdots & && & \vdots  }.
$$
But by definition $R$ is a reduced matrix, so there are no rows or columns which are all zero. Thus $a_i\neq 0$ and $b_J\neq 0$ for all $i,J$.
If $R$ {\it is separable and has a zero element}, then $a_{k^*} b_{L^*} = R_{k^*L^*} =0$; but the product of two non-zero complex number cannot be zero;
so $R$ is separable and has a zero element is contradictory. Consequently, if $R$ has a zero element, $R$ cannot be separable. In the other words, $R$ must then be entangled. $\Box$

To allow further discussion, we next define a function $\triangle$ of a matrix $M$.
\\\\
\textbf{Definition 3:}
$$
\triangle(M)\equiv
\left\{ \begin{array}{ll}
0\,\,\,\,\,\,\,\,\,\,\,\,\, if\,\,M\,\,has\,\,no\,\,zero \\
1\,\,\,\,\,\,\,\,\,\,\,\,\, otherwise.
\end{array}\right.
$$
Then \textbf{Theorem 2} can be expressed as
\begin{equation}
\triangle(R)=1 \,\,\Rightarrow\,\,R\,\,is\,\,entangled.
\end{equation}

Combining the results of all the discussions above, we finally obtain a necessary and sufficient condition for $C_{iJ}$ to be separable:
If $ | \det ( S^{(a,b)}(R)) |  = 0   $ and all $\triangle(R) =0 \Longleftrightarrow R$ has no zero elements; from \textbf{Theorem 1} we know that $R$ is then separable.
Consequently, from (\ref{eq4}), $C$ is then separable.

Conversely, if $C$ is separable, then $R$ is separable. Thus, from \textbf{Theorem 2}, it follows that $\triangle(R)=0$, namely $R$ has no zero elements. But if $R$ has no zero elements and $R$ is also separable, we can conclude that $\sum_{a,b} | \det ( S^{(a,b)}(R)) |  = 0$ from \textbf{Theorem 1}.

The net result is
\begin{equation}
\sum_{a,b} | \det ( S^{(a,b)}(R)) |  +\triangle(R)= 0\label{eq5}
\end{equation}
is a necessary and sufficient condition for $C_{iJ}$ to be separable.

\subsection{Qubit-qubit subsystems of a compound system and criterion for separability}

We would like to discuss the relation between various entanglement criteria. For this purpose we define the generalized $2\times 2$ submatrix $G $ of an  $n \times m$ matrix $C$ through  \\\\ \textbf{Definition 4:}
$$
G^{(a,b,\alpha,\beta)}(C)\equiv
\left( \begin{array}{cc}
C_{a,b}&C_{a,b+\beta}\\
C_{a+\alpha,b}&C_{a+\alpha,b+\beta}
\end{array}\right)
$$
in which the possible ranges of the elements are
$$
\left\{ \begin{array}{l}
a=1,...,(n-1),\alpha=1\\
b=1,...,(m-1),\beta=1...(m-1)
\end{array}\right.
\,\,\,\,\,\,\,or\,\,\,\,\,\,\,
\left\{ \begin{array}{l}
b=1,...,(m-1),\beta=1\\
a=1,...,(n-1),\alpha=1...(n-1).
\end{array}\right.
$$
Comparing this definition with \textbf{Definition 1}, we infer that
\begin{equation}
G^{(a,b,1,1)}=S^{(a,b)}.
\end{equation}
So we can express
\begin{equation}
\sum_{a,b,\alpha,\beta} | \det ( G^{(a,b,\alpha,\beta)}(C))|
\equiv \sum_{a,b}| \det ( S^{(a,b)}(C)) |+\chi(C).
\end{equation}
This is equivalent to defining $\chi$ by\\\\
\textbf{Definition 5:}
$$
\chi(C)\equiv \sum_{a,b,\alpha,\beta} | \det ( G^{(a,b,\alpha,\beta)}(C)) |
$$
in which
$$
\left\{ \begin{array}{l}
a=1,...,(n-1),\alpha=1\\
b=1,...,(m-2),\beta=2,...,(m-1)
\end{array}\right.
\,\,\,\,\,\,\,or\,\,\,\,\,\,\,
\left\{ \begin{array}{l}
b=1,...,(m-1),\beta=1\\
a=1,...,(n-2),\alpha=2,...,(n-1).
\end{array}\right.
$$
It follows that the following result holds.\\\\
\textbf{Theorem 3:} For a reduced matrix $R$, under condition  $\sum_{a,b} | \det ( S^{(a,b)}(R)) |  = 0$, we have
$$
\chi(R)=0 \iff \Delta(R)=0.
$$
\\
\textbf{Proof:}

From the conclusions of Eq. (\ref{eq5}), we know $ \Delta(R)=0$ implies that $R$ is separable if $\sum_{a,b} | \det ( S^{(a,b)}(R)) |  = 0 $.
If $R$ is separable, then all generalized submatrices of $R$ are separable, which implies $\chi(R)=0$.
On the other hand, if $ \Delta(R)\neq 0$, then there is at least one zero element in $R$. Assume that it is in $k^*$-th row and $L^*$-th column, namely $R_{k^*L^*}=0$.
Then, because $R$ is a reduced matrix, there must be an $u$ s.t.
\begin{equation}
(R_{k^*,u},R_{k^*,u+1})=(0,\neq 0)\,\,\,\,or\,\,\,\,(R_{k^*,u-1},R_{k^*,u})=(\neq0, 0).
\end{equation}
For the same reason, there must be a $v$ s.t. $R_{vu} \neq 0$ (because of the assumption $\sum_{a,b} | \det ( S^{(a,b)}(R)) |  = 0   $, we have $|v-k^*|\geq 2$). Therefore, there must be a generalized submatrix of $R$ which is of the form
$$
\left( \begin{array}{cc}
0&\neq 0\\
\neq 0& X
\end{array}\right),
\,\,\,\,or\,\,\,\,
\left( \begin{array}{cc}
\neq 0& X\\
0&\neq 0
\end{array}\right),
\,\,\,\,or\,\,\,\,
\left( \begin{array}{cc}
X&\neq0\\
\neq0& 0
\end{array}\right),
\,\,\,\,or\,\,\,\,
\left( \begin{array}{cc}
\neq 0&0\\
X&\neq 0
\end{array}\right).
$$
This submatrix is precisely $G^{(k^*,u,v-k^*,1)}$ for the first one.
Obviously, no matter what $X$ is, the determinant of this matrix is non-vanishing. This implies $\chi(R)\neq 0$. $\Box$

This theorem relates the following expressions.
\begin{equation}
\sum_{a,b} | \det ( S^{(a,b)}(R)) | +\chi(R) = 0 \iff \sum_{a,b} | \det ( S^{(a,b)}(R)) | + \Delta(R)=0,
\end{equation}
and
\begin{equation}
\sum_{a,b,\alpha,\beta} | \det ( G^{(a,b,\alpha,\beta)}(R)) | =0 \iff \sum_{a,b} | \det ( S^{(a,b)}(R)) | + \Delta(R)=0.
\end{equation}
So, $\sum_{a,b,\alpha,\beta} | \det ( G^{(a,b,\alpha,\beta)}(R)) | =0$ is also a necessary and sufficient condition for a reduced matrix $R$ to be separable.

From above results, we can construct our final criterion.

We define the arbitrary $2\times 2$  submatrix $Q$ of an  $n \times m$ matrix $C$ as \\\\
\textbf{Definition 6:}
$$
Q^{(s,t,u,v)}(C)\equiv
\left( \begin{array}{cc}
C_{s,u}&C_{s,v}\\
C_{t,u}&C_{t,v}
\end{array}\right),
$$
in which
$$
\left\{ \begin{array}{l}
s,t=1...n, s<t\\
u,v=1...m,u<v.
\end{array}\right.
$$
These submatrices are more general then $G$  used in the previous proof in that all qubit-qubit systems are considered and there is essentially no restriction on the ranges of the variables $(s,t,u,v)$. For this reason we adopt a different notation $Q$ instead of $G$.

If $C$ is separable,we can write $C_{iJ}=a_i b_J$. Then we have 
\begin{equation}
Q^{(s,t,u,v)}(C)=\left( \begin{array}{cc}
a_sb_u&a_sb_v\\
a_tb_u&a_tb_v
\end{array}\right).
\end{equation}
Thus $Q^{(s,t,u,v)}(C)$ is also separable and $\det(Q^{(s,t,u,v)}(C))=a_sb_ua_tb_v-a_sb_va_tb_u=0$ for all $s,t,u,v$. So we confirm
\begin{equation}
C\,\,\,is\,\,\,separable\,\,\,\Longrightarrow\,\,\,\det(Q^{(s,t,u,v)}(C))=0 \,\,\,\,\,\,\forall\,\,s,t,u,v. \label{eq101}
\end{equation}
Note that the R.H.S is equivalent to
\begin{equation}
\sum_{s,t,u,v} | \det(Q^{(s,t,u,v)}(C))| = 0.\label{eq6}
\end{equation}
Conversely, since  $Q(C)$ contains $G(R)$, when Eq. (\ref{eq6}) holds,
\begin{equation}
\sum_{a,b,\alpha,\beta} | \det ( G^{(a,b,\alpha,\beta)}(R)) | =0.
\end{equation}
So Eq. (\ref{eq6}) also implies that  $C$ is then separable.

Thus we obtain our final criterion:
\begin{equation}
\sum_{s,t,u,v} | \det(Q^{(s,t,u,v)}(C))| = 0 \iff  C \,\,\,{\rm  is \,\, separable}.\label{eq7}
\end{equation}

This revealed the interesting property that {\bf a pure bipartite composite system is separable (or non-entagled) if and only if all qubit-qubit systems of it are separable}.


\section{Entanglement parameters and their measurements}\label{sec4}
\subsection{parametrization of entanglement}

This criterion (\ref{eq7}) enables us to define numerical or quantitative measures of entanglement which {\it go beyond mere separability or non-separability determination}.
Furthermore, it reveals that for general pure states beyond qubit-qubit systems, complete accounting of entanglement can be, and has to be, described by more than one parameters.
This paves the way for finer characterizations of entanglement beyond mere separability criterion and also beyond one parameter description.

We define the entanglement parameters as follows:\\\\
\textbf{Definition 7: Entanglement parameters}
$$
E^{(s,t,u,v)} \equiv |\det(Q^{(s,t,u,v)})|^2,
$$
$$
E^{total} \equiv \sum_{s,t,u,v} E^{(s,t,u,v)}.
$$
Each $E^{(s,t,u,v)}$ represents the entanglement measure of a particular qubit-qubit subsystem $Q^{(s,t,u,v)}$, whereas $E^{total}$ represents the {\it total}  entanglement measure of the whole compound system, and it is sum of measures of entanglement of  all qubit-qubit subsystems.
Since each $E^{(s,t,u,v)}$  is positive semi-definite, our criterion (Eq. (\ref{eq7})) can be equivalently be reexpressed as
\begin{equation}
E^{total} = 0 \iff  C\quad {\rm  is\,\, separable}.
\end{equation}
Note that separability requires the vanishing of {\it each} $E^{(s,t,u,v)}$. Conversely any $E^{(s,t,u,v)} \neq 0$ implies the system is entangled.

\subsection{Measurement of entanglement parameters}

    It is shown in Ref.\cite{QQ} that the entanglement parameter of a pure bipartite qubit-qubit system is monotonic to the maximum of the expectation value of the Clauser-Horne-Shimony-Holt (CHSH) operator \cite{CHSH}\cite{KS}. This provides us a method to determine and measure the entanglement parameters described above.

Consider an arbitrary bipartite qubit-qubit system with $2\times 2$ state coefficients and state  $| \Psi \rangle =C_{iJ}{\ket i }\otimes{\ket J}$. We may label the two qubit states of each subsystem $i=\pm$ and $J=\pm$; and assume that operators $Q, R$ act on subsystem 1, while operators $S, T$ act on subsystem 2.
Then the CHSH operator can be defined as
\begin{equation}
CHSH=(Q+R) \otimes S +(Q-R)\otimes T,
\end{equation}
wherein $Q= {\hat{\vec Q}}\cdot\vec{\sigma}$ in which ${\hat{\vec Q}}$ is an arbitrary 3-dimensional unit vector and $\sigma^{i=1,2,3}$ are Pauli matrices.
Similar definition for $R, S, T$ involve other corresponding unit vectors. In Ref.\cite{QQ} it is derived and confirmed that the maximum expectation value obtained by adjusting
the vectors $({\vec Q}, {\vec R}, {\vec S}, {\vec T})$ for any {\it normalized} state $|  \Psi\rangle$ is
\begin{equation}
\left<\Psi|CHSH|  \Psi \right>_{max}=2 \sqrt{1+4|\det(C_{iJ})|^2}. \label{TT}
\end{equation}
For the case of entanglement measure which is the modulus of the determinant of the submatrix we are interested in, our submatrix coefficients $C_{iJ}$ are not necessarily normalized.
Nevertheless, we would obtained the result for non-normalized $| \Psi \rangle=C_{iJ}{\ket i }\otimes{\ket J}$ as
\begin{equation}
\left<\Psi|CHSH|  \Psi \right>_{max}=2 \sqrt{\left<\Psi|  \Psi \right>^2+4|\det(C_{iJ})|^2}\,\,\,,
\end{equation}
since substitution of $C_{iJ}$ by ${C_{iJ}}\over {\sqrt{\left<\Psi|  \Psi \right>= Tr(CC^\dagger)}}$ and ${|\Psi \rangle}$ by ${|\Psi \rangle}\over {\sqrt{\left< \Psi |\Psi \right>}}$ give the same result as \eqref{TT}  for the normalized case.

This implies that each entanglement parameter defined in our criterion can also be measured by CHSH measurements on each $2 \times 2$ submatrix which corresponds to a  particular qubit-qubit subsystem, provided $Tr(CC^\dagger)=\left<\Psi|  \Psi \right>$ for the submatrix is also known.


\section{Comparison with Peres-Horodecki criterion}\label{sec5}
\subsection{Peres-Horodecki criterion}

The Peres-Horodecki PPT (positivity of partial transpose) criterion is a well known criterion proposed by Asher Peres\cite{PER}. He proved it is necessary for separability of bipartite mixed systems. M. Horodecki, P. Horodecki and R. Horodecki further showed that the criterion is sufficient for separability for $2 \times 2$ and  $2 \times 3$ systems, but not sufficient for the case for higher-dimensional systems\cite{HOR}.

In general, a bipartite mixed system is separable iff its density matrix can be written as
\begin{equation}
\rho=\sum_i w_i \rho^A_i  \otimes  \rho_i ^B.\label{eq8}
\end{equation}
To introduce the partial transpose, Peres writes the density matrix elements explicitly with all their indices. For example, Eq. (\ref{eq8}) becomes
\begin{equation}
\rho_{m M,n N}=\sum_i w_i (\rho^A_i)_{mn}    (\rho_i ^B)_{MN}.
\end{equation}
The indices $m, n$ refer to subsystem A, and $M, N$ to subsystem B.

Thus, the definition of partial transpose is
\begin{equation}
(\rho)^{T_A}=\sigma_{m M,n N}\equiv \rho_{n M,m N},
\end{equation}
\begin{equation}
(\rho)^{T_B}=\sigma_{m M,n N}\equiv \rho_{m N,n M}.
\end{equation}
When $\rho$ is separable, Eq. (\ref{eq8}) holds, and we have
\begin{equation}
\sigma=(\rho)^{T_A}=\sum_i w_i (\rho^A_i)^T  \otimes  \rho_i ^B.
\end{equation}
The transposed matrices $(\rho^A_i)^T=(\rho^A_i)^*$ are also possible density matrices. It follows that $none \,\,of \,\,the \,\,eigenvalues \,\,of \,\, \sigma \,\,is \,\,negative$. This is a necessary condition for Eq. (\ref{eq8}) to hold, and it is the so-called Peres-Horodecki PPT (positivity of partial transpose) criterion. Showing that this criterion is also sufficient for the $2 \times 2$ and $2 \times 3$ cases is more involved. Readers can find relevant discussion in\cite{HOR}.

\subsection{Strength of Peres-Horodecki criterion for pure systems}\label{ssec52}

In this subsection we shall investigate the validity of PPT criterion for pure states in the context of our previous discussions.

First of all, we consider $2 \times 2$ case. We set
\begin{equation}
C=
\left(
\begin{array}{cc}
a&b \\
c&d
\end{array}
\right),
\end{equation}
where $a, b, c, d$ are independent variables, and satisfy the normalization condition
\begin{equation}
|a|^2+|b|^2+|c|^2+|d|^2=1
\end{equation}
Thus, these $C$ contain all possible states of a $2 \times 2$ system. The density matrix of the system is
\begin{equation}
\rho=
\left(
\begin{array}{cccc}
 a a^* & b a^* & c a^* & d a^*  \\
 a b^* & b b^* &  c b^* & d b^* \\
 a c^* & b c^*& c c^* & d c^* \\
 a d^* &b d^*& c d^* & d d^*
\end{array}
\right),
\end{equation}
and its partial transpose is
\begin{equation}
\sigma=
\left(
\begin{array}{cccc}
 a a^* & b a^* & a c^* & b c^* \\
 a b^* & b b^* & a d^* &  b d^* \\
 c a^* & d a^* & c c^* & d c^* \\
 c b^* & d b^* & c d^* & d d^*
\end{array}
\right).
\end{equation}
Through explicit calculations, we obtain the four eigenvalues of $\sigma$ which are
$$
[\pm\sqrt{b c-a d} \sqrt{b^* c^*-a^* d^*},
$$
$$
\frac{1}{2} \left(a a^*+b b^*+c c^*+d d^*\pm\sqrt{\left(a
a^*+b b^*+c c^*+d d^*\right)^2-4 (b c-a d) \left(b^* c^*-a^* d^*\right)}\right)].
$$
From this result, we can easily confirm the necessity of PPT for separability: note that $E^{total}=|\det\left(
\begin{array}{cc}
a&b \\
c&d
\end{array}
\right)|^2=|ad-bc|^2$ (=0) is necessary to guarantee PPT i.e. positivity of {\it all} the eigenvalues. And we can also confirm the sufficiency of PPT: since the term $\sqrt{b c-a d} \sqrt{b^* c^*-a^* d^*}$ appear conjugately: thus when PPT holds, this term must be zero. It follows from our discussion that $E^{total}=0$ and $C$ is separable.

We next consider $2 \times 3$ case. Similarly, we set
\begin{equation}
C=
\left(
\begin{array}{ccc}
a&b&c \\
d&e&f
\end{array}
\right),
\end{equation}
with $a, b, c, d, e, f$ satisfying the normalization condition.

Thus, the partial transpose of its density matrix is
\begin{equation}
\sigma=
\left(
\begin{array}{cccccc}
 a a^* & a b^* & a c^* & d a^* & d b^* & d c^* \\
 b a^* & b b^* & b c^* & e a^* & e b^* & e c^* \\
 c a^* & c b^* & c c^* & f a^* & f b^* & f c^* \\
 a d^* & a e^* & a f^* & d d^* & d e^* & d f^* \\
 b d^* & b e^* & b f^* & e d^* & e e^* & e f^* \\
 c d^* & c e^* & c f^* & f d^* & f e^* & f f^*
\end{array}
\right).
\end{equation}
The 6 eigenvalues are
\begin{align}
\{\notag&0,0,\\
\notag&\pm\surd[c^* \left(c d d^*-a f d^*+c e e^*-b f e^*\right)\\ \notag
&+a^* \left(-b d e^*+a e e^*-c d f^*+a f f^*\right)+b^* (b d d^*-a e d^*-c e f^*+b f f^* )],\\\notag\\\notag
&\frac{1}{2} \{a a^*+b b^*+c c^*+d d^*+e e^*+f f^* \pm\surd [(a a^*+b b^*+c c^*+d d^*+e e^*+f f^* )^2\\\notag
&+4 (c^* (-c d d^*+a f d^*-c e e^*+b f e^*)+a^* \left(b d e^*-a e e^*+c d f^*-a f f^*\right)\\\notag
&+b^* (-b d d^*+a e d^*+c e f^*-b f f^* ))]\}.
\end{align}
We can simplify these eigenvalues. With our \textbf{Definition 7} and the normalization condition, they become transparent, and are just
\begin{align}
\{\notag&0,0,\\
\notag&\pm \sqrt{E^{total}},\\\notag
&\frac{1}{2}(1\pm \sqrt{1-4E^{total}})\}
\end{align}
respectively. Thus, we deduce, similar to the $2\times2$ case, that: If $C$ is separable, $E^{total}$ equal to zero, then these eigenvalues satisfy the PPT; conversely, if we demand that PPT holds, the term $\sqrt{E^{total}}$ must be zero, and thus the system is separable (by our previous derivations). So, we confirm that the PPT criterion is also necessary and sufficient for the $2\times 3$ case.

In the same way, we calculated the eigenvalues of $\sigma$ for the $2\times4$ and $2\times5$ cases and find that there is a general rule for the form of the eigenvalues for these four cases: the eigenvalues always are
\begin{align}
\{\notag&0,0,0,... \,\,\,\,\,\,\leftarrow [{\rm (dimension \,\,of} \,\,\rho)-4 ]\,\,{\rm zeros}\\
\notag&\pm \sqrt{E^{total}},\\
&\frac{1}{2}(1\pm \sqrt{1-4E^{total}}).\}\label{eq9}
\end{align}
Consequently, for these four cases, we can draw the same conclusion that the PPT criterion is necessary and sufficient. Note that because $\sigma$ is Hermitian, the eigenvalues are all real. In Eq. (\ref{eq9}), the last two eigenvalues are $\frac{1}{2}(1\pm \sqrt{1-4E^{total}})$, and it follows that $1-4E^{total}$ cannot be negative. This gives an upper bound for $E^{total}$ for these four cases to be $\frac{1}{4}$. The maximum value of $E^{total}$ for arbitrary dimensions will be discussed in Section \ref{sec6}.

We proceed further to consider the $3 \times 3$ case. It is quite natural to ask whether Eq. (\ref{eq9}) is valid for arbitrary dimensions.
The answer is "no"; and we will see this explicitly in the $3 \times 3$ case. We likewise assume
\begin{equation}
C=
\left(
\begin{array}{ccc}
a&b&c \\
d&e&f \\
g&h&i
\end{array}
\right).
\end{equation}
The partial transpose of its density matrix is
\begin{equation}
\sigma=
\left(
\begin{array}{ccccccccc}
 a a^* & a b^* & a c^* & d a^* & d b^* & d c^* & g a^* & g b^* & g c^* \\
 b a^* & b b^* & b c^* & e a^* & e b^* & e c^* & h a^* & h b^* & h c^* \\
 c a^* & c b^* & c c^* & f a^* & f b^* & f c^* & i a^* & i b^* & i c^* \\
 a d^* & a e^* & a f^* & d d^* & d e^* & d f^* & g d^* & g e^* & g f^* \\
 b d^* & b e^* & b f^* & e d^* & e e^* & e f^* & h d^* & h e^* & h f^* \\
 c d^* & c e^* & c f^* & f d^* & f e^* & f f^* & i d^* & i e^* & i f^* \\
 a g^* & a h^* & a i^* & d g^* & d h^* & d i^* & g g^* & g h^* & g i^* \\
 b g^* & b h^* & b i^* & e g^* & e h^* & e i^* & h g^* & h h^* & h i^* \\
 c g^* & c h^* & c i^* & f g^* & f h^* & f i^* & i g^* & i h^* & i i^*
\end{array}
\right)
\end{equation}
We set $\lambda=0$, and substitute it into the eigenvalue equation of $\sigma$, to which we discover
\begin{equation}
\notag\det\left(
\begin{array}{ccccccccc}
 a a^*-\lambda  & a b^* & a c^* & d a^* & d b^* & d c^* & g a^* & g b^* & g c^* \\
 b a^* & b b^*-\lambda  & b c^* & e a^* & e b^* & e c^* & h a^* & h b^* & h c^* \\
 c a^* & c b^* & c c^*-\lambda  & f a^* & f b^* & f c^* & i a^* & i b^* & i c^* \\
 a d^* & a e^* & a f^* & d d^*-\lambda  & d e^* & d f^* & g d^* & g e^* & g f^* \\
 b d^* & b e^* & b f^* & e d^* & e e^*-\lambda  & e f^* & h d^* & h e^* & h f^* \\
 c d^* & c e^* & c f^* & f d^* & f e^* & f f^*-\lambda  & i d^* & i e^* & i f^* \\
 a g^* & a h^* & a i^* & d g^* & d h^* & d i^* & g g^*-\lambda  & g h^* & g i^* \\
 b g^* & b h^* & b i^* & e g^* & e h^* & e i^* & h g^* & h h^*-\lambda  & h i^* \\
 c g^* & c h^* & c i^* & f g^* & f h^* & f i^* & i g^* & i h^* & i i^*-\lambda
\end{array}
\right)
\end{equation}
\begin{equation}
=-|c e g-b f g-c d h+a f h+b d i-a e i|^6\neq0
\end{equation}
So, $\lambda=0$ is not an eigenvalue of $\sigma$ in general. This means that our conjecture from the previous results is wrong.

What we want to know is whether there are states which satisfy PPT but are entangled. After investigating some examples, we did not find such states. To see the strength of PPT in this case, we then solve the eigenvalue equation of $\sigma$ directly, and found of the 9 eigenvalues there are 3 eigenvalues which are solutions of the following equation:
\begin{equation}
x^3-x^2+(E^{total})x-|\det(C)|^2=0.\label{eq10}
\end{equation}
And the other six eigenvalues are $\lambda=\pm \sqrt y$, for which $y$ are solutions of the following equation:
\begin{equation}
y^3-(E^{total})y^2+|\det(C)|^2y-|\det(C)|^4=0.\label{eq11}
\end{equation}
By analyzing these two equations, we can conclude that PPT is also necessary and sufficient in this case, as will be discussed below using our previous understanding.

Necessity: We can expand $\det(C)$ as a linear combination of some submatrices Q. So, when the system is separable, the term  $(E^{total})$ and $|\det(C)|^2$ are all zero. Thus Eq. (\ref{eq10}) and Eq. (\ref{eq11}) become
\begin{equation}
x^3-x^2=0
\end{equation}
and
\begin{equation}
x^3=0\label{eq12}
\end{equation}
respectively. It follows that the eigenvalues are $(1,0,0,0,0,0,0,0,0)$, which of course satisfy the PPT criterion.

Sufficiency: Since the roots of Eq. (\ref{eq10}) are all non-negative in general, the eigenvalues corresponding to this equation all satisfy the PPT condition. When we demand that the PPT condition holds, the roots of Eq. (\ref{eq11}) must all be zero (because the eigenvalues appear conjugately). In other words, Eq. (\ref{eq11}) must be equivalent to Eq. (\ref{eq12}), that is to say, $E^{total}=|\det(C)|^2=|\det(C)|^4=0$; and thus the system is separable.

So far, we have found that PPT criterion is sufficient for many cases of pure systems, and have not found states which satisfy PPT but are entangled (even in $3\times3$ case).
This leads to the question: Is the PPT criterion sufficient for arbitrary-dimensional pure bipartite systems?


\subsection{Sufficiency of Peres-Horodecki criterion}

In this subsection we proved our conjecture using the methods and results that we have developed and obtained in this thesis.

First of all, because the trace of the density matrices of normalized is always unity and the partial transpose does not alter the value of the trace, we have
\begin{equation}
Tr( \sigma)=Tr(\rho)=1,\label{eq14}
\end{equation}
where $\rho$ is the density matrix for a bipartite pure system of arbitrary finite dimensions and $\sigma$ its partial transpose.

Secondly, we consider the trace of $\sigma^2$.  With the notation introduced earlier, we may write for pure systems,
\begin{equation}
\rho_{iJ,kL}=C_{iJ}C_{kL}^*
\end{equation}
and
\begin{equation}
\sigma_{iJ,kL}=\rho_{kJ,iL}=C_{kJ}C_{iL}^*.
\end{equation}
Thus, with these equations, we have
\begin{subequations}
\begin{align}
Tr(\sigma^2)&=\sum_{iJ,kL}\sigma_{iJ,kL} \sigma_{kLiJ}=\sum_{iJkL}C_{kJ}C_{iL}^*C_{iL}C_{kJ}^*\\
&=(\sum_{iL}|C_{iL}|^2)(\sum_{kJ}|C_{kJ}|^2)=1,\label{eq12}
\end{align}\label{eq13}
\end{subequations}
where Eq. (\ref{eq12}) follows from the normalization condition for $C$.
Furthermore, $\sigma$ is Hermitian, because of $(\sigma^\dag)_{iJ,kL}=\sigma^*_{kL,iJ}=C_{iL}^*C_{kJ}=\sigma_{iJ,kL}$. It implies that $\sigma$ is diagonalizable, and thus we have
\begin{equation}
\left\{
\begin{array}{ll}
Tr(\sigma)=\sum_i\lambda_i\\
Tr(\sigma \sigma)=\sum_i\lambda_i^2
\end{array}\right.,
\end{equation}
where $\lambda_i$ are the eigenvalues of $\sigma$. Therefore, with Eq. (\ref{eq14}) and Eq. (\ref{eq13}), we have
\begin{equation}
\sum_i\lambda_i=\sum_i\lambda_i^2=1 .\label{eq15}
\end{equation}

When we require that PPT holds, or $\lambda_i\geq0$, then Eq. (\ref{eq15}) implies
\begin{equation}
\{\lambda_i\}=\{1,0,0,0,0,...\}.
\end{equation}
This means that the rank of $ \sigma$ is one, and thus  $ \sigma$ itself is ``separable". In other words, $\exists$ $A_{iJ}$, $B_{kL}$ such that
\begin{equation}
 \sigma_{iJ,kL}=A_{iJ}B_{kL}.
\end{equation}
From our all-qubit-qubit criterion for a matrix to be separable, we know that this equation is equivalent to
\begin{equation}
\left|
\begin{array}{ll}
 \sigma_{i_1J_1,k_1L_1}& \sigma_{i_1J_1,k_2L_2} \\
 \sigma_{i_2J_2,k_1L_1}& \sigma_{i_2J_2,k_2L_2}
\end{array}
\right|=0,
\end{equation}
for all $i_1,J_1,k_1,L_1,i_2,J_2,k_2,L_2$.
Setting $J_1=J_2=J$ and $k_1=k_2=k$, we can deduce that
\begin{equation}
\left|
\begin{array}{ll}
 \sigma_{i_1J,kL_1}& \sigma_{i_1J,kL_2} \\
 \sigma_{i_2J,kL_1}& \sigma_{i_2J,kL_2}
\end{array}
\right|=0
\end{equation}
\begin{equation}\Rightarrow
\left|
\begin{array}{ll}
C_{kJ}C_{i_1L_1}^* &C_{kJ}C_{i_1L_2}^*\\
C_{kJ}C_{i_2L_1}^*&C_{kJ}C_{i_2L_2}^*
\end{array}
\right|=0
\end{equation}
\begin{equation}\Rightarrow(C_{kJ})^2
\left|
\begin{array}{ll}
C_{i_1L_1}^* &C_{i_1L_2}^*\\
C_{i_2L_1}^*&C_{i_2L_2}^*
\end{array}
\right|=0,\label{eq16}
\end{equation}
for all $i_1,i_2,L_1,L_2,J,k$. Since the matrix $C$ is normalized, it follows that $\exists k,J$ s.t. $C_{kJ}\neq0$. Then from Eq. (\ref{eq16}), we can deduce that
\begin{equation}
\left|
\begin{array}{ll}
C_{i_1L_1}^* &C_{i_1L_2}^*\\
C_{i_2L_1}^*&C_{i_2L_2}^*
\end{array}
\right|=0\,\,\,,
\end{equation}
for all $i_1,i_2,L_1,L_2$. This is equivalent to
\begin{equation}
\sum_{i_1,i_2,L_1,L_2}\left|
\begin{array}{ll}
C_{i_1L_1}^* &C_{i_1L_2}^*\\
C_{i_2L_1}^*&C_{i_2L_2}^*
\end{array}
\right|^2=0
\end{equation}
\begin{equation}\iff
E^{total}=0.
\end{equation}
And thus the system (matrix $C$) is separable. Consequently, {\it we demontrated the sufficiency of PPT for separability}.
Since Peres has proved the necessity of PPT for separability (even for mixed systems), with our result now allows us to conclude:
{\bf the Peres-Horodecki criterion is necessary and sufficient for separability of bipartite pure systems of arbitrary finite dimensions}.
This result has been confirmed in\cite{PURE} using quite different methods.

\section{Maximum value of entanglement and maximally-entangled states}\label{sec6}

In this section we shall derive the maximum value of entanglement which we defined in Section (\ref{sec4}) for arbitrary finite dimensions. We start from the $2\times m$ case. Then generalize to the $3\times m$ case, and further to the arbitrary $n\times m$ systems. In addition, the maximally-entangled states are also clarified.

\subsection{$2\times m$ case}

To simplify the indices, we express the $2\times m$ state coefficient matrix as
\begin{equation}\left(
\begin{array}{ccccc}
C_1&C_2&C_3&...&C_m \\
D_1&D_2&D_3&...&D_m
\end{array}\right).
\end{equation}
Thus we have
\begin{equation}
E^{total}=\frac{1}{2} \sum_{i \neq j}|C_iD_j-C_jD_i|^2.
\end{equation}
In subsection \ref{ssec52} we have confirmed that there is a uper bound $\frac{1}{4}$ for some $2\times m$ cases. So, here we start with the term
\begin{equation}
(\sum_i|C_i|^2+\sum_i|D_i|^2)^2-4E^{total}.
\end{equation}
It is equal to
\begin{equation}
(\sum_i|C_i|^2+\sum_i|D_i|^2)^2-4(\frac{1}{2} \sum_{i \neq j}|C_iD_j-C_jD_i|^2).
\end{equation}

We expand both these two terms, and obtain
$$
(\sum_i|C_i|^4+\sum_i|D_i|^4)+2(\frac{1}{2}\sum_{i \neq j}|C_iC_j|^2)+2(\frac{1}{2}\sum_{i \neq j}|D_iD_j|^2)
$$
$$
+2\sum_{i}|C_iD_i|^2+2\sum_{i \neq j}|C_iD_j|^2
$$
\begin{equation}
-4 \times \frac{1}{2}   \sum_{i \neq j}[ |C_iD_j|^2+|C_jD_i|^2 -C_iD_jC_j^*D_i^*-C_i^*D_j^*C_jD_i].
\end{equation}
This equals to
$$
\sum_i|C_i|^4+\sum_i|D_i|^4+2(\frac{1}{2}\sum_{i \neq j}|C_iC_j|^2)+2(\frac{1}{2}\sum_{i \neq j}|D_iD_j|^2)
$$
\begin{equation}
-2\sum_{i \neq j}|C_iD_j|^2-2\sum_{i}|C_iD_i|^2+4(\sum_{i}|C_iD_i|^2+ \sum_{i \neq j}C_i^*D_iC_jD_j^*)\,\,\,,
\end{equation}
and thus to
\begin{equation}
[\sum_{i}(|C_i|^2-|D_i|^2)]^2+4|\sum_{i}C_i^*D_i|^2.
\end{equation}
Therefore, it is obviously be non-negative, and then we have
\begin{equation}
(\sum_i|C_i|^2+\sum_i|D_i|^2)^2-4E^{total}=[\sum_{i}(|C_i|^2-|D_i|^2)]^2+4|\sum_{i}C_i^*D_i|^2\geq0\label{eq21}
\end{equation}
\begin{equation}
\Rightarrow E^{total} \leq \frac{[\sum_i|C_i|^2+\sum_i|D_i|^2]^2}{4} =\frac{1}{4} \,\,\,\label{eq17},
\end{equation}
with the normalization condition.

Thus we obtain that the maximum value of entanglement is $\frac{1}{4}$ for arbitrary $2\times m$ systems. Furthermore, from this derivation we can deduce that a state is maximally-entangled iff it is of the form
\begin{equation}\left\{
\begin{array}{l}
\sum_i|C_i|^2=\sum_i |D_i|^2 \\
\sum_{i}C_i^*D_i=0
\end{array}
\right..
\end{equation}

\subsection{$3\times m$ case}

Using the above result, we can derive the maximum value for the $3\times m$ case. The matrix of the system is expressed as
\begin{equation}\left(
\begin{array}{ccccc}
C_{11}&C_{12}&C_{13}&...&C_{1m}\\
C_{21}&C_{22}&C_{23}&...&C_{2m} \\
C_{31}&C_{32}&C_{33}&...&C_{3m}
\end{array}\right)\,\,\,,
\end{equation}
and thus the entanglement value is
\begin{subequations}
\begin{align}
E^{total}&=\sum_{i<j}\sum_{K<L}|C_{iK}C_{jL}-C_{iL}C_{jK}|^2 \\
&=\sum_{K<L}|C_{1K}C_{2L}-C_{1L}C_{2K}|^2+\sum_{K<L}|C_{2K}C_{3L}-C_{2L}C_{3K}|^2+\sum_{K<L}|C_{3K}C_{1L}-C_{3L}C_{1K}|^2\,\,,
\end{align}
\end{subequations}
where we divide it into three terms which are the entanglement ``$E^{Total}$" of the $2 \times m$ subsystem with the row indices $(1,2)$, $(2,3)$, and $(3,1)$ respectively.
It is convenient to discuss and understand in terms of other variables defined by
\begin{subequations}
\begin{align}
\alpha&\equiv |C_{11}|^2+|C_{12}|^2+|C_{13}|^2+...+|C_{1m}|^2\\
\beta&\equiv |C_{21}|^2+|C_{22}|^2+|C_{23}|^2+...+|C_{2m}|^2\\
\gamma&\equiv |C_{31}|^2+|C_{32}|^2+|C_{33}|^2+...+|C_{3m}|^2
\end{align}
\end{subequations}
and
\begin{subequations}
\begin{align}
&\mathcal{A}\equiv\alpha+\beta\\
&\mathcal{B}\equiv\beta+\gamma\\
&\mathcal{C}\equiv\gamma+\alpha.
\end{align}
\end{subequations}
Then the normalization condition can be written as
\begin{equation}
\alpha+\beta+\gamma=1.
\end{equation}

From Eq. (\ref{eq17}), we have the inequality
\begin{subequations}
\begin{align}
E^{total}&=\sum_{K<L}|C_{1K}C_{2L}-C_{1L}C_{2K}|^2+\sum_{K<L}|C_{2K}C_{3L}-C_{2L}C_{3K}|^2+\sum_{K<L}|C_{3K}C_{1L}-C_{3L}C_{1K}|^2\\
&\leq \frac{\mathcal{A}^2}{4}+\frac{\mathcal{B}^2}{4}+\frac{\mathcal{C}^2}{4}.\label{eq18}
\end{align}
\end{subequations}
The R.H.S of (\ref{eq18}) equals to
\begin{equation}
\frac{(\mathcal{A}+\mathcal{B}+\mathcal{C})^2-2(\mathcal{A}\mathcal{B}+\mathcal{B}\mathcal{C}
+\mathcal{C}\mathcal{A})}{4}
\end{equation}
\begin{equation}
=1-\frac{1}{2}(\mathcal{A}\mathcal{B}+\mathcal{B}\mathcal{C}
+\mathcal{C}\mathcal{A})\,\,\,,\label{eq20}
\end{equation}
where
\begin{equation}
\mathcal{A}\mathcal{B}+\mathcal{B}\mathcal{C}
+\mathcal{C}\mathcal{A}=\alpha^2+\beta^2+\gamma^2+3(\alpha\beta+\beta\gamma+\gamma\beta).\label{eq19}
\end{equation}
It is not a constant even with the normalization condition. But by employing an ingenious decomposition, we can still construct the inequality we need:   (\ref{eq19}) equals to
\begin{equation}
\frac{4}{3}(\alpha+\beta+\gamma)^2-\frac{1}{3}[\alpha^2+\beta^2+\gamma^2-(\alpha\beta+\beta\gamma+\gamma\alpha)].
\end{equation}
Thus Eq. (\ref{eq20}) equals to
\begin{subequations}
\begin{align}
&1-\frac{1}{2}\{\frac{4}{3}(\alpha+\beta+\gamma)^2-\frac{1}{3}[\alpha^2+\beta^2+\gamma^2-(\alpha\beta+\beta\gamma+\gamma\alpha)]\}\\
&=\frac{1}{3}+\frac{1}{6}(\frac{\alpha^2+\beta^2}{2}-\alpha\beta+\frac{\beta^2+\gamma^2}{2}-\beta\gamma+\frac{\gamma^2+\alpha^2}{2}-\gamma\alpha)\\
&=\frac{1}{3}+\frac{1}{12}[(\alpha-\beta)^2+(\beta-\gamma)^2+(\gamma-\alpha)^2].
\end{align}
\end{subequations}
And thus we have
\begin{equation}
 \frac{\mathcal{A}^2}{4}+\frac{\mathcal{B}^2}{4}+\frac{\mathcal{C}^2}{4}-\frac{1}{12}[(\alpha-\beta)^2+(\beta-\gamma)^2+(\gamma-\alpha)^2]=\frac{1}{3}.
\end{equation}
Focussing on  $\frac{1}{3}-E^{total}$, from above equations, we have
$$
\frac{1}{3}-E^{total}= \frac{\mathcal{A}^2}{4}+\frac{\mathcal{B}^2}{4}+\frac{\mathcal{C}^2}{4}-\frac{1}{12}[(\alpha-\beta)^2+(\beta-\gamma)^2+(\gamma-\alpha)^2]\,\,\,\,\,\,\,\,\,\,\,\,\,\,\,\,\,\,\,\,\,\,\,\,\,\,\,\,\,\,\,\,\,\,\,\,\,\,\,\,\,\,\,
$$
\begin{equation}
\,\,\,\,\,\,\,\,\,\,\,\,\,\,\,\,\,\,\,\,\,\,\,\,\,\,\,\,\,\,-\sum_{K<L}|C_{1K}C_{2L}-C_{1L}C_{2K}|^2-\sum_{K<L}|C_{2K}C_{3L}-C_{2L}C_{3K}|^2-\sum_{K<L}|C_{3K}C_{1L}-C_{3L}C_{1K}|^2
\end{equation}
$$
=\frac{\mathcal{A}^2}{4}-\sum_{K<L}|C_{1K}C_{2L}-C_{1L}C_{2K}|^2-\frac{1}{12}(\alpha-\beta)^2\,\,\,\,\,\,\,\,\,\,\,\,\,\,\,\,\,\,\,\,\,\,\,\,\,\,\,\,\,\,\,\,\,\,\,\,\,\,\,\,\,\,\,
$$
$$
+\frac{\mathcal{B}^2}{4}-\sum_{K<L}|C_{2K}C_{3L}-C_{2L}C_{3K}|^2-\frac{1}{12}(\beta-\gamma)^2\,\,\,\,\,\,\,\,\,\,\,\,\,\,\,\,\,\,\,\,\,\,\,\,\,\,\,\,\,\,\,\,\,\,\,\,\,\,\,\,\,
$$
\begin{equation}
+\frac{\mathcal{C}^2}{4}-\sum_{K<L}|C_{3K}C_{1L}-C_{3L}C_{1K}|^2-\frac{1}{12}(\gamma-\alpha)^2.\,\,\,\,\,\,\,\,\,\,\,\,\,\,\,\,\,\,\,\,\,\,\,\,\,\,\,\,\,\,\,\,\,\,\,\,\,\,\,\,
\end{equation}
And from Eq. (\ref{eq21}), we know this equals to
$$
\,\frac{(\alpha-\beta)^2}{4}+|\sum_iC_{1i}^*C_{2i}|^2-\frac{(\alpha-\beta)^2}{12}
$$
$$
+\frac{(\beta-\gamma)^2}{4}+|\sum_iC_{2i}^*C_{3i}|^2-\frac{(\beta-\gamma)^2}{12}
$$
\begin{equation}
+\frac{(\gamma-\alpha)^2}{4}+|\sum_iC_{3i}^*C_{1i}|^2-\frac{(\gamma-\alpha)^2}{12}
\end{equation}
$$
=\frac{(\alpha-\beta)^2}{6}+|\sum_iC_{1i}^*C_{2i}|^2\,\,\,\,\,\,\,\,\,\,\,\,\,\,\,\,\,\,\,\,\,\,\,
$$
$$
+\frac{(\beta-\gamma)^2}{6}+|\sum_iC_{2i}^*C_{3i}|^2\,\,\,\,\,\,\,\,\,\,\,\,\,\,\,\,\,\,\,\,\,
$$
\begin{equation}
+\frac{(\gamma-\alpha)^2}{6}+|\sum_iC_{3i}^*C_{1i}|^2 \,\,\,\,\,\,\,\geq 0\,\,\,.\label{eq22}
\end{equation}
Thus we obtain
\begin{equation}
E^{total}\leq \frac{1}{3}\,\,\,.
\end{equation}

This is the maximum value of entanglement for $3\times m$ systems. And from Eq. (\ref{eq22}) we infer that the state maximally-entangled iff it is of the form
\begin{equation}\left\{
\begin{array}{l}
\alpha=\beta=\gamma\\
\sum_{i}C_{1i}^*C_{2i}=\sum_{i}C_{2i}^*C_{3i}=\sum_{i}C_{3i}^*C_{1i}=0
\end{array}
\right..
\end{equation}

\subsection{$n \times m$ case}

Finally, with the previous experience, we can derive the maximum value for the $n \times m$ systems.
Although the derivation here is more complicated than in the $3\times m$ case, the ideas and steps are fortunately the same.

We express the $n \times m$ state coefficient matrix of the system as
\begin{equation}
\left(
\begin{array}{ccccc}
C_{11}&C_{12}&C_{13}&...&C_{1m}\\
C_{21}&C_{22}&C_{23}&...&C_{2m} \\
C_{31}&C_{32}&C_{33}&...&C_{3m}\\
\vdots& ...& ...&\ddots& \vdots \\
C_{n1}&C_{n2}&C_{n3}&...&C_{nm}
\end{array}\right)\,\,\,,
\end{equation}
and the total entanglement as
\begin{equation}
E^{total}=\sum_{i<j}\sum_{K<L}|C_{iK}C_{jL}-C_{iL}C_{jK}|^2.
\end{equation}
For convenience we define
\begin{equation}
L_i\equiv \sum_J |C_{iJ}|^2\,\,\,,
\end{equation}
and then the normalization condition is just
\begin{equation}
\sum_i L_i=1.
\end{equation}

Similar to the $3 \times m$ case, we regard the whole system as composed of $2 \times m$ subsystems. Thus we consider
\begin{subequations}
\begin{align}
&\sum_{i<j} \frac{(L_i+L_j)^2}{4}\\
&=\frac{[\sum_{i<j}(L_i+L_j)]^2-2[\frac{1}{2}\sum_{i<j\,,\,k<l\,,\,(i,j)\neq(k,l)}(L_i+L_j)(L_k+L_l)]}{4}\\
&=\frac{[(n-1)\sum_{i}L_i]^2-2[\frac{1}{2}\sum_{i<j\,,\,k<l\,,\,(i,j)\neq(k,l)}(L_i+L_j)(L_k+L_l)]}{4}\\
&=\frac{(n-1)^2}{4}-\frac{1}{2}[\frac{1}{2}\sum_{i<j\,,\,k<l\,,\,(i,j)\neq(k,l)}(L_i+L_j)(L_k+L_l)],
\end{align}\label{eq23}
\end{subequations}
wherein
$$
\frac{1}{2}\sum_{i<j\,,\,k<l\,,\,(i,j)\neq(k,l)}(L_i+L_j)(L_k+L_l)
$$
\begin{equation}
=\textrm{C}^{n-1}_2 (\sum_i L_i^2)+(n^2-2n)(\sum_{i<j}L_iL_j).\label{eq22}
\end{equation}
On the other hand,
\begin{equation}
(\sum_i L_i)^2=(\sum_i L_i^2)+2(\sum_{i<j}L_iL_j)=1.
\end{equation}
The question is how many times of $(\sum_i L_i^2)+2(\sum_{i<j}L_iL_j)$ should we subtract from Eq. (\ref{eq22}) so that the rest is proportional to $\sum_{i<j}(L_i-L_j)^2$. We set this multiplier to be $\mathcal {N}$. And we know that the ratio of the remaining coefficient of $(\sum_i L_i^2)$ to that of $(\sum_{i<j}L_iL_j)$ must be $(n-1): -2$; thus $\mathcal{N}$ satisfies following equation:
\begin{equation}
\textrm{C}^{n-1}_2-\mathcal{N}=-(\frac{n-1}{2})(n^2-2n-2\mathcal{N}).
\end{equation}
It follows that
\begin{equation}
\mathcal{N}=\frac{(n^2-1)(n-2)}{2n}.
\end{equation}
With this value Eq. (\ref{eq22}) equals to
\begin{subequations}
\begin{align}
&\notag \textrm{C}^{n-1}_2 (\sum_i L_i^2)+(n^2-2n)(\sum_{i<j}L_iL_j)-\mathcal{N}(\sum_i L_i)^2+\mathcal{N}(\sum_i L_i)^2\\
&=\mathcal{N}(\sum_i L_i)^2+(\textrm{C}^{n-1}_2-\mathcal{N})[(\sum_iL_i^2)-(\frac{2}{n-1})(\sum_{i<j}L_iL_j)]\\
&=\frac{(n^2-1)(n-2)}{2n}-\frac{(n-1)(n-2)}{2n}\frac{1}{(n-1)}[(n-1)(\sum_i L_i^2)-2(\sum_{i<j}L_iL_j)]\\
&=\frac{(n^2-1)(n-2)}{2n}-\frac{(n-2)}{2n}[\sum_{i<j}(L_i-L_j)^2].
\end{align}
\end{subequations}
Then Eq. (\ref{eq23}) becomes
\begin{subequations}
\begin{align}
&\notag\sum_{i<j} \frac{(L_i+L_j)^2}{4}\\
&=\frac{(n-1)^2}{4}-\frac{1}{2}[\frac{(n^2-1)(n-2)}{2n}-\frac{(n-2)}{2n}[\sum_{i<j}(L_i-L_j)^2]]\\
&=\frac{(n-1)^2}{4}-\frac{(n^2-1)(n-2)}{4n}+\frac{(n-2)}{4n}[\sum_{i<j}(L_i-L_j)^2]\\
&=\frac{n-1}{2n}+\frac{(n-2)}{4n}[\sum_{i<j}(L_i-L_j)^2].
\end{align}
\end{subequations}
Thus we have
\begin{equation}
\sum_{i<j} \frac{(L_i+L_j)^2}{4}-\frac{(n-2)}{4n}[\sum_{i<j}(L_i-L_j)^2]=\frac{n-1}{2n}.
\end{equation}
It follows that
\begin{subequations}
\begin{align}
\frac{n-1}{2n}-E^{total}&=\frac{n-1}{2n}-\sum_{i<j}\sum_{K<L}|C_{iK}C_{jL}-C_{iL}C_{jK}|^2\\
&=\sum_{i<j} [\frac{(L_i+L_j)^2}{4}-\frac{(n-2)}{4n}(L_i-L_j)^2-\sum_{K<L}|C_{iK}C_{jL}-C_{iL}C_{jK}|^2].
\end{align}
\end{subequations}
And from Eq. (\ref{eq21}) we know that this equals to
\begin{subequations}
\begin{align}
&\notag\sum_{i<j} [\frac{(L_i-L_j)^2}{4}+|\sum_KC_{iK}^*C_{jK}|^2-\frac{(n-2)}{4n}(L_i-L_j)^2]\\
&=\sum_{i<j} [\frac{1}{2n}(L_i-L_j)^2+|\sum_KC_{iK}^*C_{jK}|^2]\\
&\geq0.
\end{align}\label{eq24}
\end{subequations}
Thus we obtain an upper bound of $E^{total}$:
\begin{equation}
E^{total}\leq\frac{n-1}{2n}.
\end{equation}
And from Eq. (\ref{eq24}) we can also deduce that {\bf a generic pure bipartite $n\times m$ system is maximally-entangled states iff it is of the form}
\begin{equation}
\left\{
\begin{array}{l}
\sum_K|C_{iK}|^2=\sum_K|C_{jK}|^2\\
\sum_KC_{iK}^*C_{jK}=0
\end{array}
\right. \,\,\,\,\,for\,\,all\,\,i \neq j
\end{equation}
\begin{equation}
\iff \sum_KC_{iK}^*C_{jK}=\frac{1}{n}\delta_{ij}.
\end{equation}
This condition is equivalent to the $n$ vectors $(C_{i1},C_{i2},...,C_{im})$ being all equal in magnitude and forming an orthogonal set. If $m<n$, there cannot be as many as $n$ vectors each with $m$-components forming an orthogonal set. Then $\frac{n-1}{2n}$ is just an upper bound of the entanglement, but not the lowest upper bound in general. Therefore, strictly speaking, the maximum value of entanglement, namely the lowest upper bound, is
\begin{equation}
\frac{N-1}{2N}\,\,\,,
\end{equation}
where
\begin{equation}
N \equiv {\rm min(n,m)}.
\end{equation}

\section{Concluding remarks}

A new criterion necessary and sufficient for the separability of arbitrary finite $n \times m$ pure bipartite systems is demonstrated. It is demonstrated that the total compound system is separable if and only if all its sub-matrices are separable. This enables us to first establish the {\it necessary and sufficient criterion for separability}, and furthermore to {\it construct explicit finer quantitative measures of entanglement which go beyond mere separability or non-separability determination}. Based on this criterion, we proved that the well-known Peres-Horodecki positivity-of-partial-transform criterion is also necessary and sufficient for separability in the case of pure bipartite systems. The maximum value of entanglement, and the corresponding maximally-entangled states are also worked out in detail. We decompose the matrix of the total quantum state coefficients into  $2\times2$ sub-matrices.  The entanglement of a bipartite qubit-qubit system is monotonic to the maximum of the expectation value of the Clauser-Horne-Shimony-Holt(CHSH) operator. This implies that the finer entanglement parameters defined from our criterion can be measured by CHSH measurements on each $2\times2$ submatrix which correspond to a qubit-qubit subsystem.
Not only is the total entanglement quantifiable; our finer characterizations of entanglement determine and quantify how much entanglement there is between all possible qubit-qubit subsystems of the total system.

\begin{acknowledgments}
The author would like to thank Chopin Soo for thesis's instruction. Beneﬁcial interactions and discussions with Yu-Hsiang Chen and Hui-Chen Lin are also gratefully acknowledged.
\end{acknowledgments}


\end{document}